\documentclass[conference]{IEEEtran}
\PassOptionsToPackage{table}{xcolor}
\usepackage{graphicx}
\usepackage{multirow}
\usepackage{amsmath}
\usepackage{amssymb,soul}
\usepackage{xcolor}
\usepackage{threeparttable}
\usepackage{booktabs}
\usepackage{colortbl}
\usepackage{bbm}
\usepackage{algorithm}
\usepackage{algpseudocode}

\definecolor{rowgray}{gray}{0.92}

\ifCLASSINFOpdf

\else

\fi

\begin{document}

\title{\LARGE{Adapter-Based Few-Shot Continual Learning for Malicious Packet Recognition}}

\author{\normalsize Kyle Stein\(^1\), Guillermo Francia, III\(^2\), Eman El-Sheikh\(^2\),  Andrew Arash Mahyari\(^{1,3}\),\\
\small \(^1\) Department of Intelligent Systems and Robotics, University of West Florida, Pensacola, FL, USA\\
\small \(^2\) Center for Cybersecurity, University of West Florida, Pensacola, FL, USA\\
\small \(^3\) Florida Institute For Human and Machine Cognition (IHMC), Pensacola, FL, USA\\
\small ks209@students.uwf.edu, amahyari@ihmc.org, gfranciaiii@uwf.edu, eelsheikh@uwf.edu

}

\maketitle

\begin{abstract}
The continual evolution of malware variants necessitates detection systems that can adapt to new threats without retraining from scratch. However, continually updating models on new data often leads to catastrophic forgetting, where previously learned knowledge is overwritten. While continual learning has been increasingly explored for malware detection, the specific setting of Few-Shot Class-Incremental Learning (FSCIL), where new malware classes must be learned from only a small number of labeled examples, remains comparatively underexplored. Therefore, this work investigates the FSCIL setting for malware classification. To address the stability-plasticity dilemma, we propose a hybrid framework that leverages a Self-Supervised Learning (SSL) backbone initialized through domain-specific pre-training on malware packets. Our method incorporates Low-Rank Adaptation (LoRA) to efficiently adapt the model during the base session while freezing the core backbone to preserve previously learned representations, alongside a prototype-based classification head for incremental sessions to establish robust decision boundaries from limited samples. Extensive experiments across several datasets demonstrate that our approach consistently outperforms prior malware FSCIL baselines and achieves state-of-the-art performance.

%The continual evolution of malware variants necessitates detection systems that can adapt to new threats without retraining from scratch. However, continually updating models on new data often leads to catastrophic forgetting, where previously learned knowledge is overwritten. While continual learning addresses this challenge, the specific application of Few-Shot Class-Incremental Learning (FSCIL) to malware detection remains underexplored. Existing research often assumes abundant data for new classes, lacking the depth required to handle emerging threats where samples are scarce. Therefore, this work conducts an investigation specifically into the FSCIL scenario applied to malware classification. To address the stability-plasticity dilemma in FSCIL, we propose a hybrid framework leveraging a Self-Supervised Learning (SSL) backbone initialized with domain-specific pre-training on malware packets. Our method incorporates Low-Rank Adaptation (LoRA) to efficiently fine-tune the robust base session while freezing the core backbone to preserve stability, alongside a prototype-based classification head for incremental sessions to establish robust decision boundaries. Extensive experiments on several datasets demonstrate that our approach consistently outperforms prior malware FSCIL baselines and achieves state-of-the-art performance.
\end{abstract}


\section{Introduction}
\label{sec:introduction}

Deep learning has transformed cybersecurity, enabling the detection and classification of malicious network traffic with unprecedented accuracy \cite{de2021machine, stein2024revolutionizing, deeppacket}. However, deployed malware classifiers face a fundamental challenge in the real world when the threat landscape continuously evolves, with new malware families emerging faster than large labeled datasets can be curated \cite{tayyab2022survey}. A classifier trained on known threats must possess the ability to rapidly adapt to novel attack patterns without full retraining of the model. In our previous work \cite{stein2024towards}, we showed that a detector can adapt to previously unseen malware families in a static few-shot setting, where the set of novel target classes is fixed. In contrast, modern deployments face a sequential, evolving scenario in which new malware families may arrive in multiple sessions over time and the model must incorporate them while retaining performance on all previously learned classes, a setting that naturally results in catastrophic forgetting \cite{french1999catastrophic}. The field of continual learning, or class-incremental learning (CIL), formalizes this setting where a learner must adapt to new classes over time without forgetting previously learned ones \cite{de2021continual, zhou2024continual}. 

Traditional CIL scenarios consist of balanced sessions, where new classes are introduced over time, and where each new class has many samples each to learn from. Many CIL methods have been proposed to address the issue of forgetting, mainly in computer vision and image recognition tasks \cite{zhou2024continual, he2022exemplar, zhou2024expandable}. However, a more recent field of few-shot class incremental learning (FSCIL) \cite{tao2020few} was formulated to address catastrophic forgetting in a new environment: where it is assumed that a model, trained on a robust base session of classes, must adapt to new classes with only few samples for each new class. In both of these scenarios, it is intuitive to rely on rehearsal \cite{rebuffi2017icarl}, where the model replays previously seen samples while learning new classes to preserve earlier knowledge and stabilize the feature space of incremental sessions. 

While rehearsal strategies have traditionally been effective at mitigating forgetting, they introduce significant deployment concerns. In many real-world scenarios, privacy regulations mandate that data be discarded immediately after processing. This constraint is particularly critical in malware analysis, where maintaining a persistent buffer of live malicious code poses a severe security risk, potentially leading to accidental execution or leakage \cite{guri2019using, or2019dynamic}. Consequently, a rehearsal-free framework is essential for secure continual learning, as it allows the system to discard malware samples immediately after model updates. However, the absence of a replay buffer typically hurts the stability and plasticity needed for continual learning models. Without past data to anchor the optimization, updating the model typically degrades the feature space and collapses previously learned decision boundaries \cite{zhou2024class}. %Therefore, effective malware adaptation requires rehearsal-free mechanisms that can accommodate novel threats without overwriting the weights essential for recognizing established families.

In this work, we propose a multi-faceted approach to rehearsal-free FSCIL for malware detection through a modern, adapter-based strategy. Motivated by recent successes in leveraging pre-trained models for rehearsal-free CIL \cite{zhou2024class, zhou2024continual, park2024pre}, we first pre-train a transformer from scratch on malware packets using a self-supervised objective \cite{devlin2018bert}. We then adapt the pre-trained backbone to incremental sessions with low-rank adaptation (LoRA) \cite{hu2022lora} and perform classification using a prototype-based decision rule. Despite its simplicity, our approach outperforms previous continual learning methods for malware classification that rely on complex architectures. Finally, we evaluate under a fine-grained protocol where specific attack subtypes are treated as independent classes. This approach avoids simplifying the task through hierarchical grouping, thereby forcing the model to resolve high inter-class similarity and substantially increasing the difficulty of incremental adaptation.

Overall, our contributions can be summarized as follows:

\begin{itemize}
    \item We introduce domain-specific self-supervised pre-training on malware payloads with parameter-efficient LoRA updates on a frozen backbone and a prototype-based classifier to preserve decision boundaries over time.
    \item We propose the first framework to combine self-supervised pre-training from scratch with adapter-based few-shot continual learning for malicious packet recognition, enabling efficient adaptation to emerging malware classes without retaining previously observed samples.
    \item We provide an extensive empirical evaluation across multiple benchmark datasets, achieving state-of-the-art performance in few-shot continual malware classification.
\end{itemize}

\section{Related Work}
\label{sec:related-work}

\subsection{Few-Shot Class Incremental Learning and Rehearsal}
FSCIL \cite{tao2020few} extends CIL \cite{zhou2024class, liu2025class} by adapting to novel classes with scarce labeled data. While earlier approaches relied on metric learning \cite{yang2023neural} or full fine-tuning \cite{peng2022few}, recent trends leverage frozen pre-trained transformers adapted through Parameter-Efficient Fine-Tuning (PEFT). Common prompt-based PEFT methods \cite{wang2022learning, smith2023coda} often face optimization difficulties \cite{ding2023parameter}. To address this, we utilize an adapter-based strategy \cite{zhou2024expandable, he2025cl}, injecting lightweight modules into the frozen backbone to mitigate forgetting without the slow convergence associated with prompting.

Rehearsal, or experience replay \cite{rebuffi2017icarl}, mitigates forgetting by storing and replaying past exemplars. While effective at maintaining decision boundaries, storing raw malware samples poses severe security and privacy risks, such as accidental execution or leakage \cite{thomas2017ethical, botes2022cryptographic}. Consequently, the field is shifting toward rehearsal-free methods \cite{smith2023coda, he2025cl} that rely on pre-trained models rather than stored data. In this work, we explicitly evaluate the performance trade-off between rehearsal-based and rehearsal-free architectures specifically within the malware domain.

\subsection{Continual Learning for Malware Classification}
An early finding in continual learning for malware classification is that many continual learning techniques, originally developed and benchmarked in computer vision, do not transfer cleanly to malware. Rahman \textit{et al.} \cite{rahman2022limitations} evaluated a broad set of CL methods on malware datasets and found that many approaches fail to prevent catastrophic forgetting. This motivates malware-specific investigations into which continual learning assumptions are actually realistic for security deployments. FSCIL malware detection has also been explored in \cite{chai2024malfscil}, where a base-stage model is trained with a replay-like variational autoencoder mechanism, and incremental sessions freeze the feature extractor while updating a prototype-based classifier. BFS-NID \cite{du2023few} mitigates forgetting for FSCIL by fixing the feature extractor parameters after the base session and utilizing a branch classifier learning module. In contrast to prior work, we propose the first malicious packet recognition framework to combine domain-specific self-supervised pre-training from scratch with adapter-based FSCIL, enabling rehearsal-free adaptation under scarce incremental data.

\section{Preliminaries}
\label{sec:preliminaries}

% \subsection{CIL Problem Setup} 
% In Class-Incremental Learning (CIL), the model incrementally learns to recognize a set of distinct semantic categories from a global label set $\mathcal{Y}=\{y_1, y_2, \dots, y_K\}$. Learning proceeds over a sequence of $T$ tasks $\mathcal{T} = \{\mathcal{T}_0, \mathcal{T}_1, \dots, \mathcal{T}_{T-1}\}$, where each task $\mathcal{T}_t$ introduces a disjoint set of classes $\mathcal{Y}^{(t)} \subset \mathcal{Y}$, such that $\mathcal{Y}^{(i)} \cap \mathcal{Y}^{(j)} = \emptyset$ for any $i \neq j$. Consequently, the model is exposed to new classes at each step without any overlap with previously encountered classes. Each task $\mathcal{T}_t$ provides a labeled dataset $\mathcal{D}^{(t)} = \{(x_i, y_i)\}$, where each label $y_i \in \mathcal{Y}^{(t)}$ denotes the specific class category of the input $x_i$. The model must learn to recognize the new classes while maintaining performance on all previously seen classes in the cumulative label set $\mathcal{Y}^{(\le t)} = \bigcup_{i=0}^t \mathcal{Y}^{(i)}$.

\begin{figure*}[!htbp]
\centering
\includegraphics[width=1.0\linewidth]{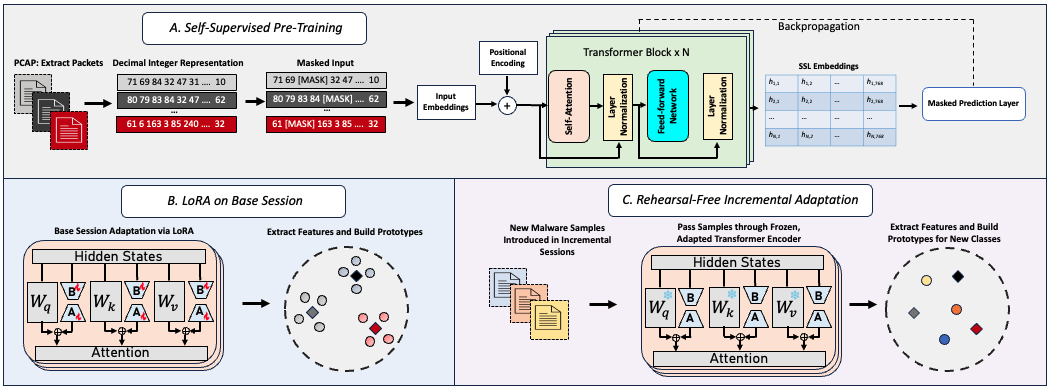}
\caption{The overall architecture of the proposed rehearsal-free FSCIL approach for malicious packet recognition.} 
\label{fig:Main_Fig}
\end{figure*}

\subsection{FSCIL Problem Setup} 
FSCIL follows a similar sequential structure to CIL, but introduces severe data constraints for later tasks. Learning proceeds over a sequence of $T$ tasks $\mathcal{T} = \{\mathcal{T}_0, \mathcal{T}_1, \dots, \mathcal{T}_{T-1}\}$, where the first task $\mathcal{T}_0$ is denoted as the \textit{base session}. In the base session, a comprehensive label space $\mathcal{Y}^{(0)}$ is provided along with ample data for each class, establishing a robust feature representation. However, for each subsequent incremental session $\mathcal{T}_t$ (where $1 \le t < T$), only a few training samples per class are available. These incremental datasets $\mathcal{D}^{(t)}$ are arranged in an $N$-way $K$-shot format, where $N$ new classes are introduced with only $K$ labeled examples each. As in CIL, the class sets are disjoint: $\mathcal{Y}^{(i)} \cap \mathcal{Y}^{(j)} = \emptyset$ for any $i \neq j$. During inference after session $t$, the model is evaluated on a query test set $\mathcal{D}^{(\leq t)}_{\text{test}}$ containing samples from the cumulative label set $\mathcal{Y}^{(\leq t)} = \bigcup_{j=0}^{t} \mathcal{Y}^{(j)}$. The goal is to incorporate new classes from scarce examples without catastrophic forgetting of the previously learned classes.

\section{Proposed Method}
\label{sec:proposed-method}
In this section, we describe the architecture and training protocol of our rehearsal-free malicious packet recognition FSCIL framework. Our approach operates in three distinct stages: self-supervised pre-training of a byte-level transformer encoder to learn protocol semantics from scratch, PEFT on base classes using LoRA, and a rehearsal-free incremental prototype classification stage for novel malware families. The overall framework of our approach is shown in Fig. \ref{fig:Main_Fig}.

\subsection{Transformer Encoder}
Adapting pre-trained models to downstream tasks has gained significant traction due to the ability of large-scale models to be efficiently fine-tuned for specialized classification tasks \cite{zhou2024class, han2024parameter}. This is prevalent in computer vision, where transformers are routinely fine-tuned for continual object classification. Motivated by these insights, we propose self-supervised pre-training of a transformer backbone on raw malware packets to learn robust, domain-specific feature representations before the introduction of class labels, shown to be effective in our previous work \cite{stein2024revolutionizing}. We employ a transformer encoder \cite{devlin2018bert} to process network packets as sequences of bytes. Unlike traditional Natural Language Processing (NLP) which operates on word tokens, our model treats each byte value (0-255) as a discrete token. To handle variable-length inputs and masking, we extend the byte token space size to $V=258$, reserving indices 256 for padding and 257 for masking.

The backbone consists of an embedding layer, positional encodings, and a stack of multi-head self-attention blocks. Given an input packet sequence $\mathbf{X} = [x_1, \dots, x_L]$, the embedding layer maps each byte to a vector of dimension $d_{model}=768$. Standard positional encodings are added to preserve sequence order information. The self-attention mechanism computes contextual dependencies between bytes through:

\begin{equation}
\text{Attention}(\mathbf{Q}, \mathbf{K}, \mathbf{V}) = \text{softmax}\left(\frac{\mathbf{Q}\mathbf{K}^\top}{\sqrt{d_k}}\right)\mathbf{V},
\end{equation}

\noindent where $\mathbf{Q}, \mathbf{K}, \mathbf{V}$ are the query, key, and value projections. We utilize a mean-pooling operation over the final hidden states to extract a fixed-size global representation vector $\mathbf{e} \in \mathbb{R}^{d_{model}}$ for each packet.

\subsection{Self-Supervised Pre-training}
To learn robust packet representations without explicit labeled classes, we train the backbone using a Masked Language Modeling (MLM) objective. Given an unlabelled packet sequence $\mathbf{x}$, we generate a binary mask $M \in \{0,1\}^L$ where each position has a 15\% probability of being masked. This forces the model to reconstruct the original byte value $x_i$ solely relying on the contextual information provided by the surrounding unmasked bytes.

The model projects the output hidden states of masked positions to the byte token space size $V$ using a linear prediction head. We optimize the parameters $\theta$ by minimizing the cross-entropy loss between the predicted probabilities and the original byte indices:

\begin{equation}
\mathcal{L}_{MLM} = -\sum_{i \in \mathcal{M}} \log P(x_i | \tilde{\mathbf{x}}; \theta),
\end{equation}

\noindent where $\mathcal{M}$ is the set of masked indices and $\tilde{\mathbf{x}}$ represents the masked input sequence. We utilize the AdamW optimizer with a learning rate of $1\times10^{-4}$ and weight decay to mitigate overfitting during this phase. This yields a general-purpose feature extractor $f_\theta(\cdot)$ trained on malicious packet bytes. To adapt this extractor to specific malware families without destroying these general features, we proceed with a PEFT strategy with LoRA on the base session.

\subsection{Base Session Adaptation with LoRA}
Following pre-training, full fine-tuning on labeled data can lead to catastrophic forgetting of the pre-trained features due to overfitting to the few, newly introduced samples \cite{park2024pre}. Therefore, it is important to develop a system that retains knowledge learned during the base session while remaining robust to the few malware samples introduced in later incremental sessions. However, this is a difficult task since we must learn from only $K$ malware samples in incremental sessions due to our FSCIL constraints. To address this, we adopt an adapter-based approach through LoRA \cite{hu2022lora}, which enables efficient adaptation of pre-trained models by freezing the weights learned during pre-training and injecting trainable low-rank matrices into the attention layers. 

For a pre-trained weight matrix $\mathbf{W}_0 \in \mathbb{R}^{d_{\text{out}} \times d_{\text{in}}}$, LoRA modifies the forward pass as:

\begin{equation}
\mathbf{y} = \mathbf{W}_0\mathbf{x} + \Delta\mathbf{W}\mathbf{x},
\label{eq:lora}
\end{equation}

\noindent where $\mathbf{x}$ and $\mathbf{y}$ denote the input and output activations, respectively. The update $\Delta\mathbf{W} = \mathbf{A}\mathbf{B}$ is decomposed into low-rank matrices $\mathbf{B} \in \mathbb{R}^{r \times d_{\text{in}}}$ and $\mathbf{A} \in \mathbb{R}^{d_{\text{out}} \times r}$, where $r \ll \min(d_{\text{in}}, d_{\text{out}})$ is a user-chosen rank controlling the capacity of the adaptation. In this paper, we set the rank to 8 for all experiments. By constraining the update to this low-rank structure, LoRA captures the most important task-specific changes while requiring far fewer trainable parameters than updating the full weight matrix. We apply LoRA to the query, key, and value projections of our transformer-based packet encoder. %For all subsequent incremental sessions, the adapters are permanently frozen since unfreezing the adapter may lead to severe catastrophic forgetting in attempting to learn from the newly introduced few samples. %Unlike standard continual learning approaches that train separate adapters for every task, we optimize the LoRA parameters solely during the base session. 

\subsection{Rehearsal-Free Incremental Learning}
While LoRA enables efficient adaptation in the base session, continuing to update the model during incremental few-shot sessions can cause feature drift and degrade previously learned decision boundaries. However, a frozen feature space is viable in our setting because malware packet classes share underlying structural semantics. Our prior work demonstrated that different malware classes share such semantics in their packets \cite{stein2024revolutionizing, stein2024towards}. By leveraging SSL during pre-training and LoRA during the base session, we establish a universal feature extractor that does not require further gradient updates to recognize new classes. As previously discussed, reliance on replay buffers would violate rehearsal-free constraints because it requires storing privacy-sensitive historical data. Therefore, we design a lightweight prototype construction phase that simply averages the available features of the newly introduced classes.

\vspace{1mm}
\noindent \textbf{Prototype Formation:} Formally, when $K$-shot support samples are provided for a new class $c$, we compute its prototype $\mathbf{p}_c$ by averaging the $L_2$-normalized feature vectors of the support set $S_c$, followed by a re-normalization step:

\begin{equation}
\mathbf{p}_c = \text{Normalize}\left( \frac{1}{|S_c|} \sum_{\mathbf{x} \in S_c} \frac{f_\theta(\mathbf{x})}{\|f_\theta(\mathbf{x})\|_2} \right)
\end{equation}

\noindent where $f_\theta$ represents the frozen, pre-trained feature extractor and $\text{Normalize}(\mathbf{v}) = \mathbf{v} / \|\mathbf{v}\|_2$. 

\vspace{1mm}
\noindent \textbf{Inference:} 
Unlike traditional Prototypical Networks \cite{snell2017prototypical} that utilize Euclidean distance, we employ a scaled cosine-similarity classifier to align with the normalized embedding space. Given the set of class prototypes $\{\mathbf{p}_c\}$ and a query embedding $\mathbf{z}_q = \text{Normalize}(f_\theta(\mathbf{x}_q))$, we compute the similarity scores:

\vspace{-3mm}
\begin{equation}
s_c = \gamma \cdot (\mathbf{z}_q^\top \mathbf{p}_c)
\end{equation}
\vspace{-3mm}

\noindent where $\gamma$ is a scaling factor and $s_c$ represents the logit for class $c$. These logits are converted into class probabilities via the softmax function:

\vspace{-3mm}
\begin{equation}
p(y=c \mid \mathbf{x}_q) = \frac{\exp(s_c)}{\sum_{j} \exp(s_j)}.
\end{equation}
\vspace{-3mm}

\noindent The predicted label for $\mathbf{x}_q$ is $\arg\max_c\, p(y=c \mid \mathbf{x}_q)$. Since $f_\theta(\cdot)$ remains frozen (no gradients are back-propagated) during our incremental testing scenario, we perform inference directly using these computed similarities.

\subsection{Rehearsal-Based Incremental Learning}
To provide a comparative baseline, we implement a rehearsal-based variant that relaxes standard privacy constraints to store a small buffer of exemplars. Unlike our proposed method, which relies on a static feature space, this variant continues parameter updates during incremental sessions using a combined dataset of the current few-shot samples and stored exemplars from previously seen classes. The replayed exemplar buffer is mixed with the new samples during training to constrain feature drift and reduce forgetting. After each session, we update the memory bank with exemplars from the newly introduced classes and refresh the prototype set by re-encoding the stored exemplars under the updated model to recompute class centroids. We argue that this strategy is suboptimal since it requires storing sensitive raw payloads, and the overfitting induced by repeated updates on limited data can degrade the generalizability of the SSL and base-session features.

\begin{table}[t!]
\centering
\scriptsize 
\caption{Experimental Splits. Both benchmarks ($S_0$ Base, $S_{1-3}$ Incremental) are evaluated under identical 5-shot constraints with 70 query samples per incremental class.}
\label{tab:combined_sessions}
\resizebox{\columnwidth}{!}{%
\begin{tabular}{c|c|c|l|l}
\hline
\multirow{2}{*}{\textbf{Sess.}} & \multirow{2}{*}{\textbf{Task Type}} & \multirow{2}{*}{\textbf{\begin{tabular}[c]{@{}c@{}}Samples\\ (Train / Test)\end{tabular}}} & \multicolumn{2}{c}{\textbf{Attack Class Composition}} \\ \cline{4-5} 
 &  &  & \textbf{CIC-IDS2017} & \textbf{UNSW-NB15} \\ \hline
\multirow{4}{*}{$S_0$} & \multirow{4}{*}{\begin{tabular}[c]{@{}c@{}}Base\\ (Many-Shot)\end{tabular}} & \multirow{4}{*}{500 / 200} & $\bullet$ DDoS & $\bullet$ Generic \\
 &  &  & $\bullet$ Infiltration & $\bullet$ Analysis \\
 &  &  & $\bullet$ DoS Slowhttptest & $\bullet$ Reconnaissance \\
 &  &  & $\bullet$ FTP-Patator & \textit{---} \\ \hline
\multirow{3}{*}{$S_1$} & \multirow{3}{*}{\begin{tabular}[c]{@{}c@{}}Inc.\\ (5-Shot)\end{tabular}} & \multirow{3}{*}{5 / 70} & $\bullet$ SSH-Patator & $\bullet$ Shellcode \\
 &  &  & $\bullet$ Heartbleed & $\bullet$ DoS \\
 &  &  & $\bullet$ DoS Slowloris & \textit{---} \\ \hline
\multirow{3}{*}{$S_2$} & \multirow{3}{*}{\begin{tabular}[c]{@{}c@{}}Inc.\\ (5-Shot)\end{tabular}} & \multirow{3}{*}{5 / 70} & $\bullet$ Web – Brute Force & $\bullet$ Backdoor \\
 &  &  & $\bullet$ Web – XSS & $\bullet$ Fuzzers \\
 &  &  & $\bullet$ DoS GoldenEye & \textit{---} \\ \hline
\multirow{3}{*}{$S_3$} & \multirow{3}{*}{\begin{tabular}[c]{@{}c@{}}Inc.\\ (5-Shot)\end{tabular}} & \multirow{3}{*}{5 / 70} & $\bullet$ Bot & $\bullet$ Exploits \\
 &  &  & $\bullet$ PortScan & $\bullet$ Worms \\
 &  &  & $\bullet$ DoS Hulk & \textit{---} \\ \hline
\end{tabular}%
}
\end{table}

\begin{table*}[t]
\centering
\caption{5-Shot FSCIL Performance on UNSW-NB15 and CIC-IDS2017.}
\label{tab:fscil-main}
\begin{threeparttable}
\renewcommand{\arraystretch}{1.2}
\setlength{\tabcolsep}{6pt} 
\begin{tabular}{l l l c c c c c | c}
\toprule
\textbf{Dataset} & \textbf{Strategy} & \textbf{Method} & \textbf{\# Exemplars} & \textbf{S0} & \textbf{S1} & \textbf{S2} & \textbf{S3} & \textbf{FGT} \\
\midrule

\multirow{6}{*}{\textbf{CIC-IDS2017}} 
 & \multirow{3}{*}{No-Rehearsal} 
   & BFS-NID \cite{du2023few} & 0 & $98.20$\text{\scriptsize$\pm0.86$} & $75.91$\text{\scriptsize$\pm3.13$} & $64.61$\text{\scriptsize$\pm1.46$} & $52.30$\text{\scriptsize$\pm5.07$} & $22.05$\text{\scriptsize$\pm3.98$} \\
 & & MalFSCIL \cite{chai2024malfscil} & 0 & $84.80$\text{\scriptsize$\pm1.53$} & $65.11$\text{\scriptsize$\pm2.09$} & $52.92$\text{\scriptsize$\pm2.38$} & $41.90$\text{\scriptsize$\pm2.46$} & $17.28$\text{\scriptsize$\pm4.98$} \\
 & & \textbf{Proposed} & 0 & $\mathbf{99.92}$\text{\scriptsize$\mathbf{\pm0.70}$} & $\mathbf{83.79}$\text{\scriptsize$\mathbf{\pm1.69}$} & $\mathbf{71.01}$\text{\scriptsize$\mathbf{\pm2.87}$} & $\mathbf{61.05}$\text{\scriptsize$\mathbf{\pm2.47}$} & $\mathbf{8.59}$\text{\scriptsize$\mathbf{\pm1.54}$} \\
 \cmidrule{2-9}
 
 & \multirow{2}{*}{Rehearsal} 
 & Proposed & 1 & $99.88$\text{\scriptsize$\pm0.10$} & $76.77$\text{\scriptsize$\pm3.08$} & $58.30$\text{\scriptsize$\pm5.09$} & $53.23$\text{\scriptsize$\pm6.58$} & $29.20$\text{\scriptsize$\pm8.19$} \\
 & & \textbf{Proposed} & 5 & $\mathbf{99.91}$\text{\scriptsize$\mathbf{\pm0.11}$} & $\mathbf{84.46}$\text{\scriptsize$\mathbf{\pm3.43}$} & $\mathbf{70.46}$\text{\scriptsize$\mathbf{\pm3.74}$} & $\mathbf{59.54}$\text{\scriptsize$\mathbf{\pm5.68}$} & $\mathbf{19.95}$\text{\scriptsize$\mathbf{\pm2.13}$} \\

\midrule
\midrule

\multirow{6}{*}{\textbf{UNSW-NB15}} 
 & \multirow{3}{*}{No-Rehearsal} 
   & BFS-NID \cite{du2023few} & 0 & $83.66$\text{\scriptsize$\pm0.83$} & $74.19$\text{\scriptsize$\pm1.04$} & $60.50$\text{\scriptsize$\pm2.62$} & $50.44$\text{\scriptsize$\pm6.81$} & $13.67$\text{\scriptsize$\pm5.48$} \\
 & & MalFSCIL \cite{chai2024malfscil} & 0 & $75.94$\text{\scriptsize$\pm12.59$} & $51.22$\text{\scriptsize$\pm16.55$} & $23.18$\text{\scriptsize$\pm6.77$} & $17.01$\text{\scriptsize$\pm4.79$} & $34.91$\text{\scriptsize$\pm4.63$} \\
 & & \textbf{Proposed} & 0 & $\mathbf{88.56}$\text{\scriptsize$\mathbf{\pm0.42}$} & $\mathbf{74.77}$\text{\scriptsize$\mathbf{\pm0.88}$} & $\mathbf{64.66}$\text{\scriptsize$\pm\mathbf{4.36}$} & $\mathbf{58.59}$\text{\scriptsize$\mathbf{\pm5.38}$} & $\mathbf{9.14}$\text{\scriptsize$\mathbf{\pm4.70}$} \\
 \cmidrule{2-9} 
 
 & \multirow{2}{*}{Rehearsal} 
 & Proposed & 1 & $85.39$\text{\scriptsize$\pm0.89$} & $61.53$\text{\scriptsize$\pm12.90$} & $47.77$\text{\scriptsize$\pm9.59$} & $40.66$\text{\scriptsize$\pm2.62$} & $26.85$\text{\scriptsize$\pm6.42$} \\
 & & \textbf{Proposed} & 5 & $\mathbf{88.89}$\text{\scriptsize$\mathbf{\pm1.15}$} & $\mathbf{74.73}$\text{\scriptsize$\mathbf{\pm2.92}$} & $\mathbf{53.67}$\text{\scriptsize$\mathbf{\pm2.32}$} & $\mathbf{50.01}$\text{\scriptsize$\mathbf{\pm3.27}$} & $\mathbf{17.76}$\text{\scriptsize$\mathbf{\pm1.33}$} \\

\bottomrule
\end{tabular}
\footnotesize
\end{threeparttable}
\end{table*}

\section{Experimental Results}
\label{sec:results}
\noindent \textbf{Datasets and Metrics.} 
Our method is evaluated on two renowned malicious datasets, specifically CIC-IDS2017 \cite{sharafaldin2018toward} and UNSW-NB15 \cite{UNSW}, utilizing the extracted datasets made available through Payload-byte \cite{farrukh2022payload}. Table \ref{tab:combined_sessions} details the dataset splits across both benchmarks. Furthermore, SSL pre-training is performed using the Session 0 split, which is used to learn the pre-trained encoder initialization. Consistent with standard FSCIL protocols \cite{tao2020few}, we divide the training process into a robust base session ($S_0$) and sequential incremental sessions ($S_1, \dots, S_N$). For the base session, we assume sufficient data availability and utilize 500 training samples per class to establish a discriminative initial feature space. For all subsequent incremental sessions, we enforce a N-way, K-Shot format, where the model must learn new attack patterns using only $K$ support samples. 

We evaluate on a cumulative test set containing query samples from all classes observed up to the current session. All results are averaged over three random seeds. We employ two standard metrics: \textbf{Accuracy} on the cumulative test set at the current session and \textbf{Average Forgetting} measures the decline in knowledge retention. For each previously learned class, we calculate the difference between \textit{peak accuracy} (the highest accuracy ever achieved for the class in any prior session) and the \textit{current accuracy}. We report the average of these drops across all old classes. All experiments are performed on a single NVIDIA H100 GPU.

\subsection{Main Experimental Results}
In this section, we discuss our main experimental results presented in Table \ref{tab:fscil-main}. On both CIC-IDS2017 and UNSW-NB15, our proposed rehearsal-free method significantly outperforms existing baselines. Against the SOA Rehearsal-Free baseline methods, our method achieves a final session ($S_3$) accuracy improvement of $+8.75\%$ on CIC-IDS2017 ($61.05\%$ vs $52.30\%$) and $+8.15\%$ on UNSW-NB15 ($58.59\%$ vs $50.44\%$). Furthermore, our method demonstrates superior stability, reducing the average forgetting rate by roughly half compared to MalFSCIL. This indicates that the combination of SSL pre-training and LoRA adapters establishes a far more robust feature space than the variational autoencoder approach used in prior work.

A key observation in our results is the performance trade-off between our Rehearsal-Based (Buffer=5) and Rehearsal-Free (Buffer=0) variants. As shown in Table \ref{tab:fscil-main}, both methods achieve comparable global accuracy in the final session (e.g., $61.05\%$ vs $59.54\%$ on CIC-IDS2017). However, their underlying behavior differs fundamentally regarding average forgetting. The rehearsal-based variant exhibits significantly higher forgetting ($19.95\%$ on CIC-IDS2017) compared to the Rehearsal-Free approach ($8.59\%$). This illustrates the classic stability-plasticity dilemma. The rehearsal-based method allows for plasticity (updating LoRA weights), which enables it to learn new incremental classes effectively, but this weight update causes a drift in the feature space, leading to the forgetting of base classes. However, our rehearsal-free method enforces strict stability by freezing the weights. While the incremental sesssions remain static, the robust SSL-initialized backbone ensures that the base classes are preserved with minimal degradation. The rehearsal-free method achieves superior or equivalent overall performance without the security risks associated with storing a replay buffer, validating our hypothesis that a static, robustly pre-trained feature space is preferable for malware FSCIL.

\section{Limitations}
While this work advances the state of FSCIL for malware detection, it is important to discuss the limitations associated with this field. First, our approach relies on the self-supervised pre-training of a transformer backbone from scratch which incurs significant initial computational costs. Although improvements in GPUs and hardware makes this manageable, and the use of SSL eliminates the need for labeled data during this phase, the quality of the resulting model remains heavily dependent on the diversity of the unlabeled malware corpus used for pre-training. Next, our rehearsal-free strategy enforces a static feature space during incremental sessions to prevent catastrophic forgetting. While effective for stability, this design assumes that the pre-trained backbone and base-session adapters yield sufficiently discriminative features for all future malware variants. If novel attacks exhibit a significant domain shift relative to the pre-training distribution, the frozen encoder may limit plasticity compared to methods that continuously fine-tune the backbone. However, we empirically address this concern in our experiments by injecting diverse classes during incremental sessions, thereby demonstrating the robustness of our feature space to significant semantic shifts. Finally, the proposed framework is specifically optimized for dynamic threat landscapes where retraining is prohibitive. In static environments where the set of malware families changes infrequently, the complexity of a continual learning system may exceed that of simpler, static classifiers.

% \begin{table}[t]
% \centering
% \scriptsize
% \caption{Ablation on LoRA rank $r$ for the proposed method. Final session accuracy (S3) and forgetting (FGT) are reported.}
% \label{tab:lora_rank_ablation}
% \begin{tabular}{c cc cc}
% \toprule
% \multirow{2}{*}{\textbf{Rank $r$}} & \multicolumn{2}{c}{\textbf{CIC-IDS2017}} & \multicolumn{2}{c}{\textbf{UNSW-NB15}} \\
% \cmidrule(lr){2-3} \cmidrule(lr){4-5}
%  & \textbf{S3} & \textbf{FGT} & \textbf{S3} & \textbf{FGT} \\
% \midrule
% 2  &  &  &  &  \\
% 4  &  & & 55.38 & 6.18 \\
% 8  & 61.05 & 8.59 & 51.19 & 11.97 \\
% 16 &  &  &  &  \\
% \bottomrule
% \end{tabular}
% \end{table}

\section{Conclusion}
\label{sec:conclusion}
In this paper, we proposed a rehearsal-free FSCIL framework for malware classification that leverages SSL pre-training and LoRA to address catastrophic forgetting. By selectively updating only the self-attention layers from the domain specific SSL objective, our method effectively integrates new classes from scarce samples while preserving robust base representations. Extensive experiments on CIC-IDS2017 and UNSW-NB15 demonstrate that our approach significantly outperforms previous baselines in stability, showing that a robust frozen feature space is preferable to data replay in security-sensitive domains. Our findings underscore the advantages of targeted parameter tuning for deploying adaptive intrusion detection systems where data retention is prohibited.

\vspace{-2mm}
\section{Acknowledgment}
This work is partially supported by the UWF Argo Cyber Emerging Scholars (ACES) program funded by the National Science Foundation (NSF) CyberCorps\textsuperscript{\textregistered} Scholarship for Service (SFS) award under grant number 1946442. Any opinions, findings, and conclusions or recommendations expressed in this document are those of the authors and do not necessarily reflect the views of the NSF.

\ifCLASSOPTIONcaptionsoff
  \newpage
\fi

\bibliographystyle{IEEEtran}% or another style like plain, alpha, etc.
\bibliography{ref}

\end{document}